\documentclass[useAMS, usenatbib]{mn2e}

\usepackage[dvips]{graphicx}
\input{epsf}

\title{Ghosts of the Milky Way: a search for topology in new quasar catalogues}
\author[S. J. Weatherley, et al.]
       {S.~J.~Weatherley\thanks{Email : stephen.weatherley@imperial.ac.uk}$^1$, S.~J.~Warren$^1$, S.~M.~Croom$^2$, R.~J.~Smith$^3$, B.~J.~Boyle$^2$, \newauthor T.~Shanks$^4$, L.~Miller$^5$ and M.~P.~Baltovic$^6$\\
$^1$Astrophysics Group, Blackett Laboratory, Imperial College London, Prince Consort Road, London SW7 2BW, UK\\
$^2$Anglo-Australian Observatory, PO Box 296, Epping, NSW 1710, Australia\\
$^3$Astrophysics Research Institute, Liverpool John Moores University, Twelve Quays House, Egerton Wharf, Birkenhead CH41 1LD, UK\\
$^4$Department of Physics, University of Durham, South Road, Durham DH1 3LE, UK\\
$^5$Department of Physics, Oxford University, 1 Keeble Road, Oxford OX1 3RH, UK\\
$^6$Theoretical Physics Group, Blackett Laboratory, Imperial College London, Prince Consort Road, London SW7 2BW, UK}
\date{Accepted 0000 January 00.
      Received 0000 January 00;
      in original form 0000 January 00}

\pagerange{\pageref{firstpage}--\pageref{lastpage}}
\pubyear{2003}
\begin{document}

\maketitle

\label{firstpage}
%%%%%%%%%%%%%%%%%%%%%%%%%%%%%%%%%%%%%%%%%%%%%%%%%%%%%%%%%%%%%%%%
%%%%%%%%%%%%%%%%%%%%%%%%%%%%%%%%%%%%%%%%%%%%%%%%%%%%%%%%%%%%%%%%
\begin{abstract}

We revisit the possibility that we inhabit a compact multi-connected
flat, or nearly-flat, Universe. Analysis of {\it COBE} data has shown that, 
for such a case, the size of the fundamental domain must be a substantial fraction of
the horizon size. Nevertheless, there could be several copies of the
Universe within the horizon. If the Milky Way was once a quasar we
might detect its `ghost' images. Using new large quasar catalogues we
repeat the search by Fagundes \& Wichoski for antipodal quasar
pairs. By applying linear theory to account for the peculiar velocity
of the local group, we are able to narrow the search radius to
$134$ arcsec. We find seven candidate antipodal quasar pairs within
this search radius. However, a similar number would be expected by
chance. We argue that, even with larger quasar catalogues, and more
accurate values of the cosmological parameters, it is unlikely to be
possible to identify putative ghost pairs unambiguously, because of
the uncertainty of the correction for peculiar motion of the Milky
Way.

\end{abstract}

%%%%%%%%%%%%%%%%%%%%%%%%%%%%%%%%%%%%%%%%%%%%%%%%%%%%%%%%%%%%%%%
%%%%%%%%%%%%%%%%%%%%%%%%%%%%%%%%%%%%%%%%%%%%%%%%%%%%%%%%%%%%%%%


%%%%%%%%%%%%%%%%%%%%%%%%%%%%%%%%%%%%%%%%%%%%%%%%%%%%%%%%%%%%%%%
%%%%%%%%%%%%%%%%%%%%%%%%%%%%%%%%%%%%%%%%%%%%%%%%%%%%%%%%%%%%%%%

\section{Introduction}

The idea that we inhabit a compact multi-connected Universe has
been presented by many authors (see review by \citet{LRL}). 
The simplest such Universe is one in which space
is flat and may be visualised as tiled by a `fundamental parallelepiped' with
lengths $\alpha_x$, $\alpha_y$, $\alpha_z$. If the sides of the
fundamental domain are significantly smaller than the horizon length,
R$_H$, observable effects would be present. One of the most
interesting, albeit surreal, effects would be the appearance of `ghost
images'. This is where radiation from an object, such as a galaxy, has
traversed the fundamental domain, resulting in a repeat or ghost image
of the object. The simplest ghost images to search for are ones of our
own galaxy, since they appear at the nodes of a lattice of which the
Milky Way is the origin.  Therefore ghost images might be most simply identified
by searching for antipodal pairs. In this paper we merge three new
large quasar catalogues to search for ghost images of the Milky Way.

The best limits on the size of the fundamental domain come from
analyses of the {\it  COBE} measurement of the cosmic microwave
background. The simplest space is the hypertorus. This is a spatially flat topology 
where opposite faces
of the parallelepiped are identified. There are three other compact
spaces, involving twists of $\pi/2$ or $\pi$ in pairing opposite
faces of a parallelepiped (e.g. \citealt{L1998}). There are also two compact models 
in which the fundamental polyhedron is a hexagonal prism with twists of $\pi/3$ and $2\pi/3$ (\citealt{LRL}). 
For equal sides of the parallelepiped, \citet{L1998} obtained a minimum size of the
side of the fundamental domain of 0.8R$_H$, and a similar result for the hexagonal prism.
 \citet{Bond} obtained a somewhat larger limit for the hypertorus. For unequal sides the
constraints are less strong, and there remains the possibility that
there are many copies of the Universe within the horizon. The existence of antipodal
pairs is a common feature of most topologies whether the geometry is hyperbolic,
flat or spherical. To this extent, therefore, although we assume a flat geometry our 
analysis is also relevant to nearly-flat geometries.

On the basis of these {\it COBE} limits, it is clear that any search for ghosts
needs to use objects at large redshifts, and so catalogues of quasars
are an obvious source. Many authors have used quasars to probe
topological structure (\citealt{F1987}; \citealt{F1989};
\citealt{R1996}), but have yet to constrain topology using these
methods. Following the suggestion that many large galaxies go through
a quasar phase, \citet{F1987} searched a quasar catalogue for
antipodal quasar pairs in the hope of finding ghost images of the
Milky Way. In fact even with complete catalogues of quasars over the
whole sky, good fortune would be required for success with such a
search.  This is because quasar lifetimes must be short relative to
the age of the Universe, so the method relies on the quasar being
alive at the particular time corresponding to the look-back time of
any copies. Nevertheless quasar catalogues have grown enormously in
the last few years, due in particular to the contributions of the
2-degree-field QSO survey (hereafter 2dF) and the Sloan Digital Sky
Survey (hereafter SDSS). Therefore we considered it timely to repeat
the search of Fagundes \& Wichoski for antipodal quasar pairs.

The outline of the paper is as follows. In \S2 we describe the quasar
catalogues used. In \S3 we explain the search method. The search is
complicated by the peculiar velocity of the Milky Way $v_{\circ}$,
which displaces the ghost images from opposition by order
$v_{\circ}/c$, and so must be accounted for. A suitable search radius,
then, depends on the uncertainty of the historical motion of the Milky
Way, which depends, in turn, on the uncertainty of the cosmological
density parameter. The results of the search are presented in \S4, together
with a brief discussion. We assume flat Universal geometry $(\Omega_{\rm m} 
+ \Omega_\Lambda = 1)$ throughout and consider only topologies without axial twists.

%%%%%%%%%%%%%%%%%%%%%%%%%%%%%%%%%%%%%%%%%%%%%%%%%%%%%%%%%%%%%%%%
%%%%%%%%%%%%%%%%%%%%%%%%%%%%%%%%%%%%%%%%%%%%%%%%%%%%%%%%%%%%%%%%

\section{Quasar Catalogues}
The quasar catalogue used for this investigation was compiled from
three sources:
\begin{enumerate}

\item The complete candidate list for the 2dF QSO survey (Smith et
al., 2003, in prep.). This catalogue contains 47,768
objects, from which we extracted the 23,340 quasars with spectroscopic
redshifts. A subset of 10,000 of the
quasars has been published in \citet{Croom}. 

\item The early data release from the Sloan Digital Sky Survey Quasar
Catalogue \citep{SDSS} which contains 3,814 quasars.

\item The quasar catalogue of \citet{VV}\footnote{\scriptsize available at:
\textsf{http://www.obs-hp.fr/www/catalogues/veron2\_10/veron2\_10.html}}
which contains 23,760 quasars.

\end{enumerate}

\begin{figure}
\includegraphics[width=1\columnwidth]{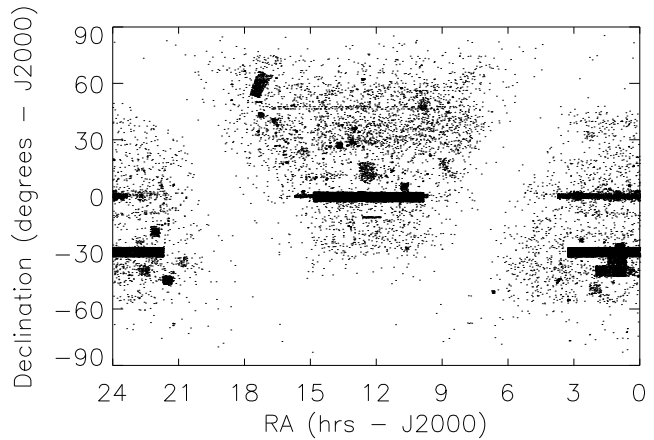}
\caption{Sky distribution of the 43,271 quasars in the merged catalogue
used for the search. \label{figure:quasars}}
\end{figure}

Duplicate quasars, primarily the 10,000 quasars in \citet{Croom}
 which are included in the V\'eron catalogue, need to be
eliminated. We searched the merged list for pairs of objects within
2 arcsec and $\Delta z < 0.02$, and in each case eliminated
one of the pair. This could, in fact, have included real pairs, but
this would not affect our search for antipodal pairs, since it uses a
much larger search radius. This left a final catalogue of 43,271
objects reaching a maximum redshift of $z=5.8$. The sky distribution
of the quasars in this merged catalogue is illustrated in Figure
\ref{figure:quasars}.

%%%%%%%%%%%%%%%%%%%%%%%%%%%%%%%%%%%%%%%%%%%%%%%%%%%%%%%%%%%%%%%%%%%
%%%%%%%%%%%%%%%%%%%%%%%%%%%%%%%%%%%%%%%%%%%%%%%%%%%%%%%%%%%%%%%%%%%

\section{Method}

\subsection{Peculiar Velocity}

Because of the past motion of the local group, ghost images of the
Milky Way will appear displaced from the lattice nodes. This angular
displacement is of order $v_{\circ}/c\sim 7$ arcmin, and must be
corrected for. The effect is greatest for objects lying perpendicular
to the direction of local group motion, and falls to zero for objects
that lie on the path of local group motion. Therefore the idea is to
apply an appropriate angular correction for each object in our
catalogue to bring true antipodal ghost images into opposition. The
uncertainty of the computed correction sets the search radius, since
it is much greater than the uncertainty of the quasar coordinates, or
any other relevant effect. Previous searches for antipodal quasar
pairs have failed to correct for the motion of the local group. 

To find this angular correction we turn to linear theory for an
expression for peculiar velocity. From this we can find the comoving
distance moved by the local group as a function of redshift, and from this
compute the appropriate angular correction for each quasar.

The peculiar velocity, $v_{pec}$, is related to the peculiar
acceleration, $\bmath{g}$, by
\[
\bmath{v}_{pec} = \frac{a(z)}{a(0)}\bmath{\dot{x}} = \frac{2f}{3\Omega H}\bmath{g}
\]
\citep{P1993} where $a(z)$ is the scale factor at epoch $z$ and
$\bmath{\dot{x}}$ is a comoving velocity. Here
\[
\Omega = \Omega_{\rm m} \big(\frac{H}{H_{\circ}}\big)^{-2}(1+z)^3.
\]
The term $f$ is the dimensionless
velocity factor, given by
\[
f = \frac{d \log \delta}{d \log a}
\]
where $\delta$ is the overdensity or density contrast. For a flat
universe, a good approximation to $f$ is given by
\[
f = \Omega^{0.6} + \frac{1}{70}\big[1 - \frac{1}{2}\Omega(1 + \Omega)\big]
\]
\citep{Lahav}.

In our linear approximation the peculiar acceleration is given by 
\[
\bmath{g}=\frac{GM_{p}(\bmath{x}_{p}-\bmath{x}_{lg})}{a^2|\bmath{x}_{p}-\bmath{x}_{lg}|^3}
\]
where $\bmath{x}$ is a comoving distance and the subscripts refer to a
single point-attractor, {\it p}, and the local group, {\it lg}. This
equation provides the scaling of $\bmath{g}$ with $a$, but the mass of
the attractor, and its distance, are not needed explicitly, since
we have a measurement of the peculiar velocity today.

The expression for peculiar velocity is numerically integrated to find
the comoving distance moved by the local group as a function of
redshift. This is then converted to an angular correction for a ghost
image at redshift $z$. The only input variables to this correction
algorithm are $\Omega_{\rm m}$ and $\bmath{v}_{\circ}$. The values adopted for
this analysis are:
\begin{enumerate}
\item $\Omega_{\rm m} = 0.3 \pm 0.1$  (e.g. \citealt{Verde}). 
\item $\bmath{v}_{\circ} = 627 \pm 22$km\,s$^{-1}$ in direction $(l,b) = (276\degr \pm 3\degr, 30\degr \pm 3\degr)$ from {\it COBE}
data \citep{Kogut}.
\end{enumerate}

\begin{figure}
\includegraphics[width=1\columnwidth]{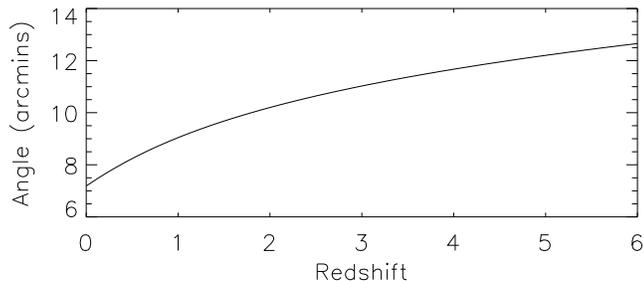}
\caption{The angular correction to the quasar position as a function
of redshift, for a quasar located in a direction perpendicular to the
direction of local group motion (maximum
correction). \label{figure:angle}}
\end{figure}

The angular correction for each object depends on redshift and the
angle between the quasar, the observer, and the direction of local
group motion. For an object located perpendicular to the direction of
local group motion (maximum correction) this correction is
approximately 7 arcmin at $z=0$, and increases to approximately 12
arcmin at $z=5$. A plot of this angular correction is shown in Figure
\ref{figure:angle}.

%%%%%%%%%%%%%%%%%%%%%%%%%%%%%%%%%%%%%%%%%%%%%%%%%%%%%%%%%%%%%%%%%%%%%%%
%%%%%%%%%%%%%%%%%%%%%%%%%%%%%%%%%%%%%%%%%%%%%%%%%%%%%%%%%%%%%%%%%%%%%%%

\subsection{Search Radius}
An appropriate search radius must be related to the uncertainties in
the input data, and also involves a trade-off between contamination
and incompleteness. Too large a tolerance and we recover too many
false positives, swamping any real pairs, while with too small a
tolerance we risk losing the few putative real pairs in the data.

We required that the redshifts of the pair agree within a tolerance
$\Delta z=0.04$. The large majority of the redshifts in the catalogue
have been measured from intermediate resolution spectra, many by cross
correlation against a template spectrum. For these the accuracy of the
redshift is better than $\Delta z=0.01$ (e.g. \citealt{Croom}). 
Therefore the chosen tolerance will include nearly all real
pairs in the catalogue.

The position of any quasar in the catalogue, after correction for the
peculiar velocity of the local group, is uncertain, because of the
uncertainty of the parameters $\Omega_{\rm m}$ and
$\bmath{v}_{\circ}$. Gravitational lensing could also contribute,
since the light paths from antipodal quasars traverse different
regions of the fundamental domain. This is a much smaller source of
uncertainty, however, and may be neglected. It would be very unlikely
for a ghost pair to be lensed out of opposition by more than a few
arcsec.

In principle we could compute a probability distribution function
(pdf) for the coordinates of each quasar, and select a contour level
at, say, the $95$ per cent confidence level. Then, in searching for an
antipodal quasar, the size of the contour should be doubled. This is
because the error in the peculiar velocity correction (due to poor
choice of $\Omega_{\rm m}$, $\bmath{v}_{\circ}$) is identical for each
quasar of an antipodal pair. This procedure would be computationally
expensive, and we followed a simpler Monte Carlo approach. The
algorithm to correct the coordinates was run on the catalogue 200
times, each time using values of the input parameters
$\Omega_{\rm m}$, $\bmath{v}_{\circ}$, chosen from a Gaussian
distribution about their mean. We computed the angular displacement
from the mean position for each quasar in every simulation, and
selected the angle which included $95$ per cent of all the points. This angle
was $67$ arcsec. Therefore the procedure for the search was to
correct the quasar coordinates using the best estimates for
$\Omega_{\rm m}$, and $\bmath{v}_{\circ}$, and then for each quasar to
search for antipodal quasars in the catalogue within a search angular
radius of $134$ arcsec. Because the coordinate pdfs for each quasar
differ in size, shape, and orientation, in some cases this search
radius will be unnecessarily large, and in some cases too small. But,
by construction, it will include $95$ per cent of any real antipodal pairs.

We found that the largest source of error is from the uncertainties in
our current knowledge of the speed and direction of the local group
relative to the CMB. Our motion {\it within} the local group is a
source of error unaccounted for in our search radius determination. As
we are searching for a ghost of the Galaxy rather than a ghost of the
local group, our motion within it should be considered. However, this
motion is small compared with the large scale streaming motion of the
local group itself.

%%%%%%%%%%%%%%%%%%%%%%%%%%%%%%%%%%%%%%%%%%%%%%%%%%%%%%%%%%%%%%%%%%%%%%
%%%%%%%%%%%%%%%%%%%%%%%%%%%%%%%%%%%%%%%%%%%%%%%%%%%%%%%%%%%%%%%%%%%%%%

\section{Results and discussion}
With a search radius of $134$ arcsec and $\Delta z=0.04$ we recover
seven antipodal quasars pairs. Details of the pairs are provided in
Table \ref{table:quasars}. The table includes the coordinates of each
quasar in the pair both before and after correction for peculiar velocity as well as
the redshifts, brightnesses, the catalogue containing the quasar,
 the redshift difference, and the angular displacement from opposition
of each pair. 

\begin{figure*}
\includegraphics[width=5.5cm]{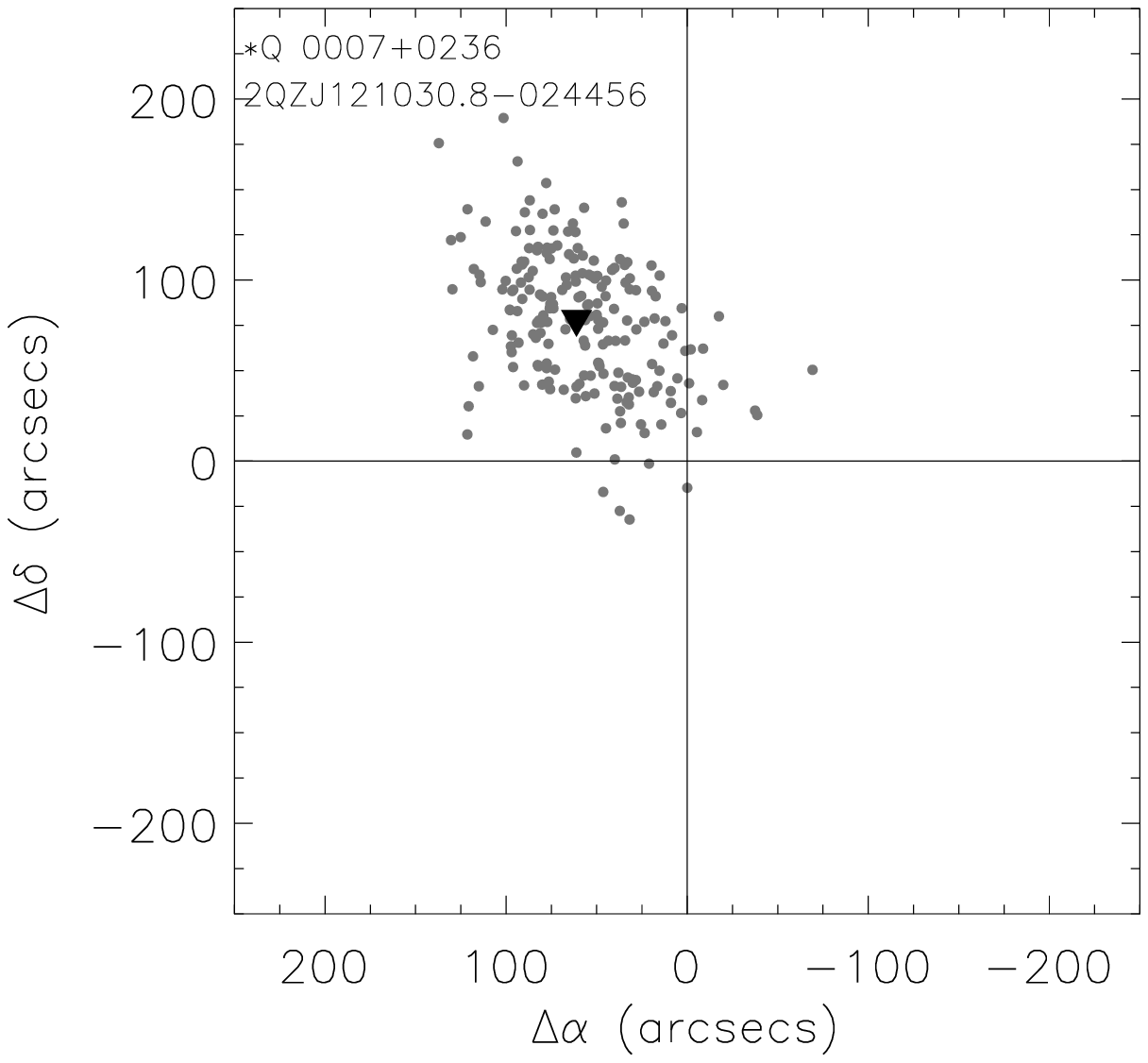}
\includegraphics[width=5.5cm]{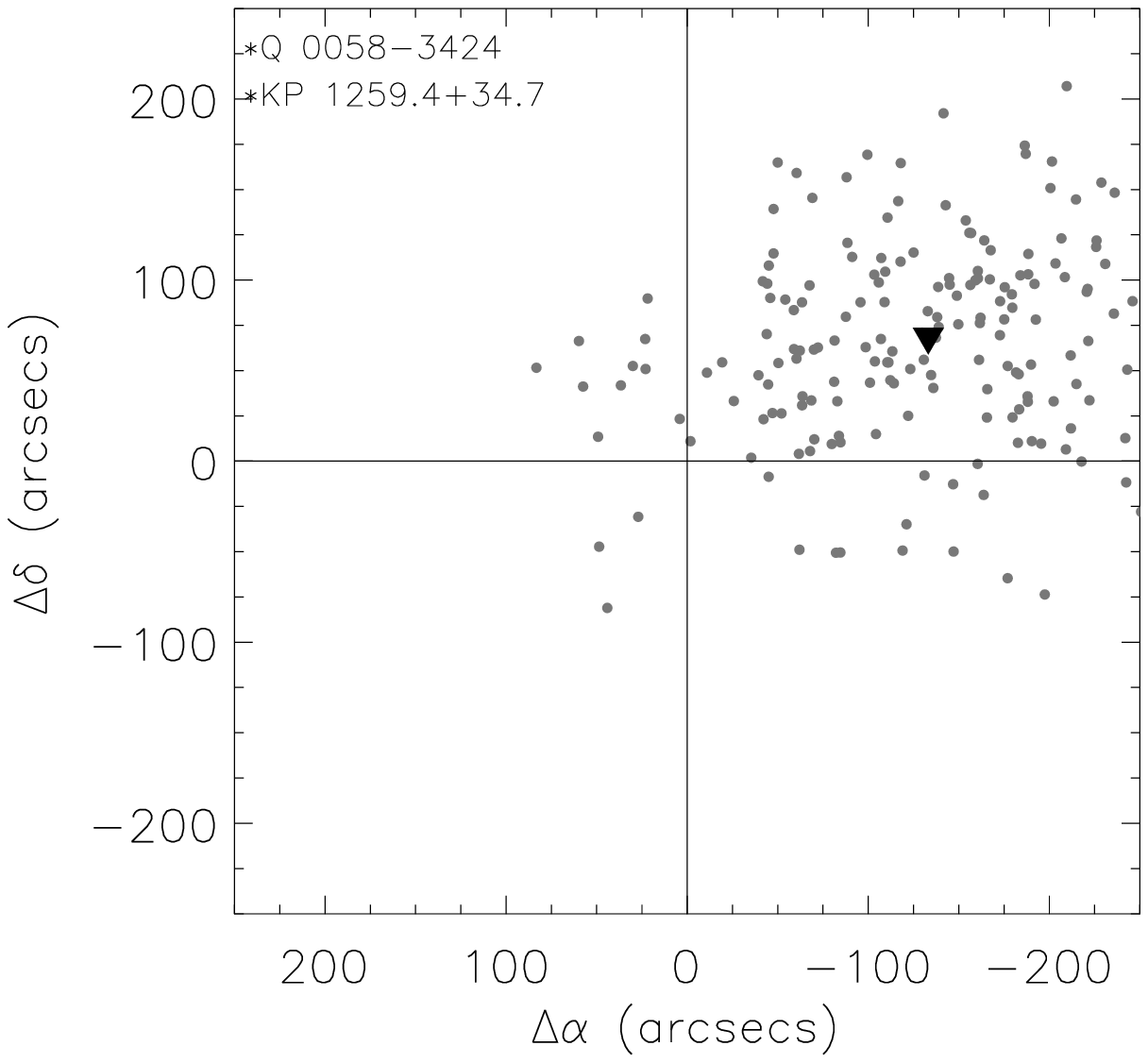}
\includegraphics[width=5.5cm]{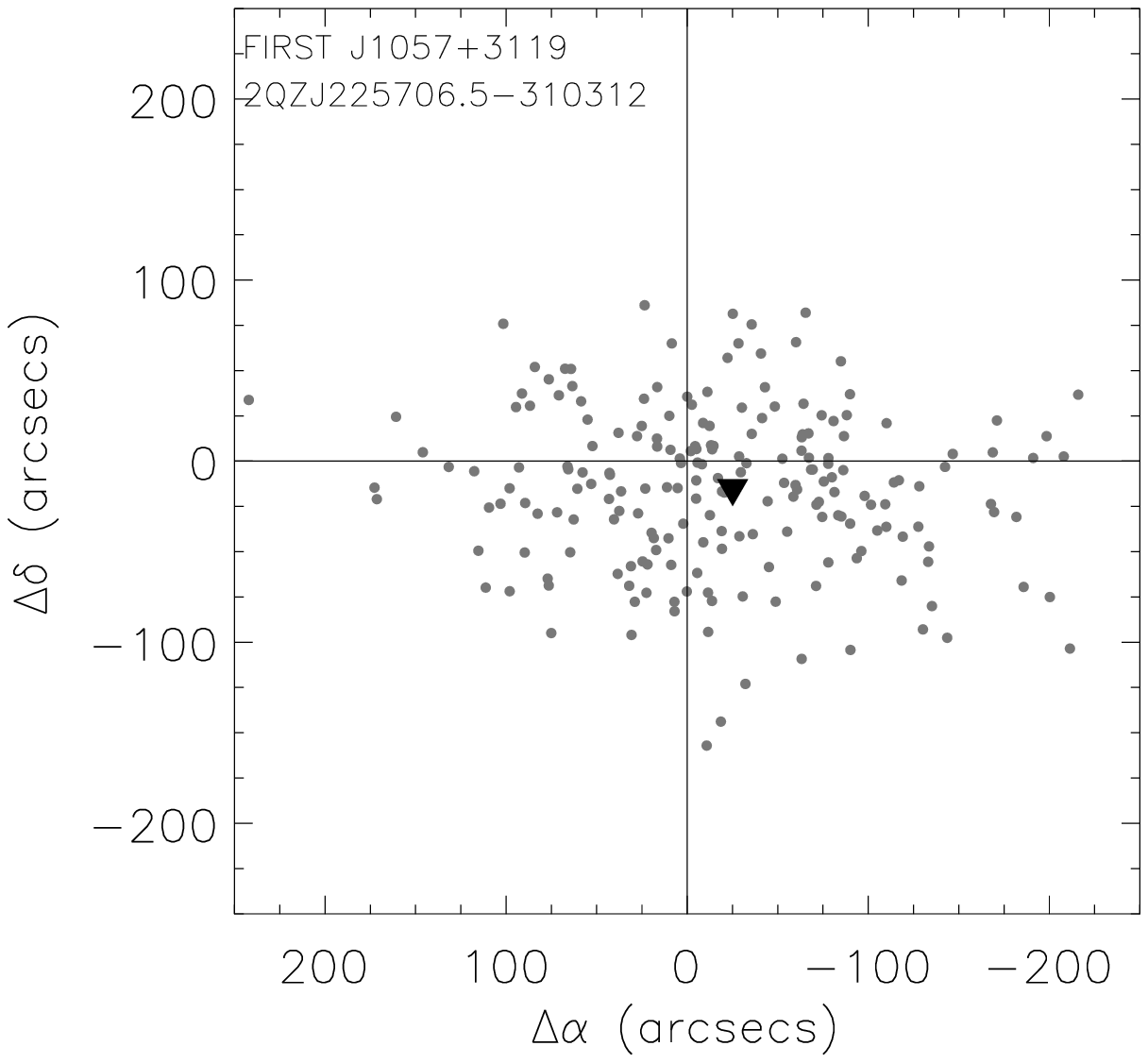}
\includegraphics[width=5.5cm]{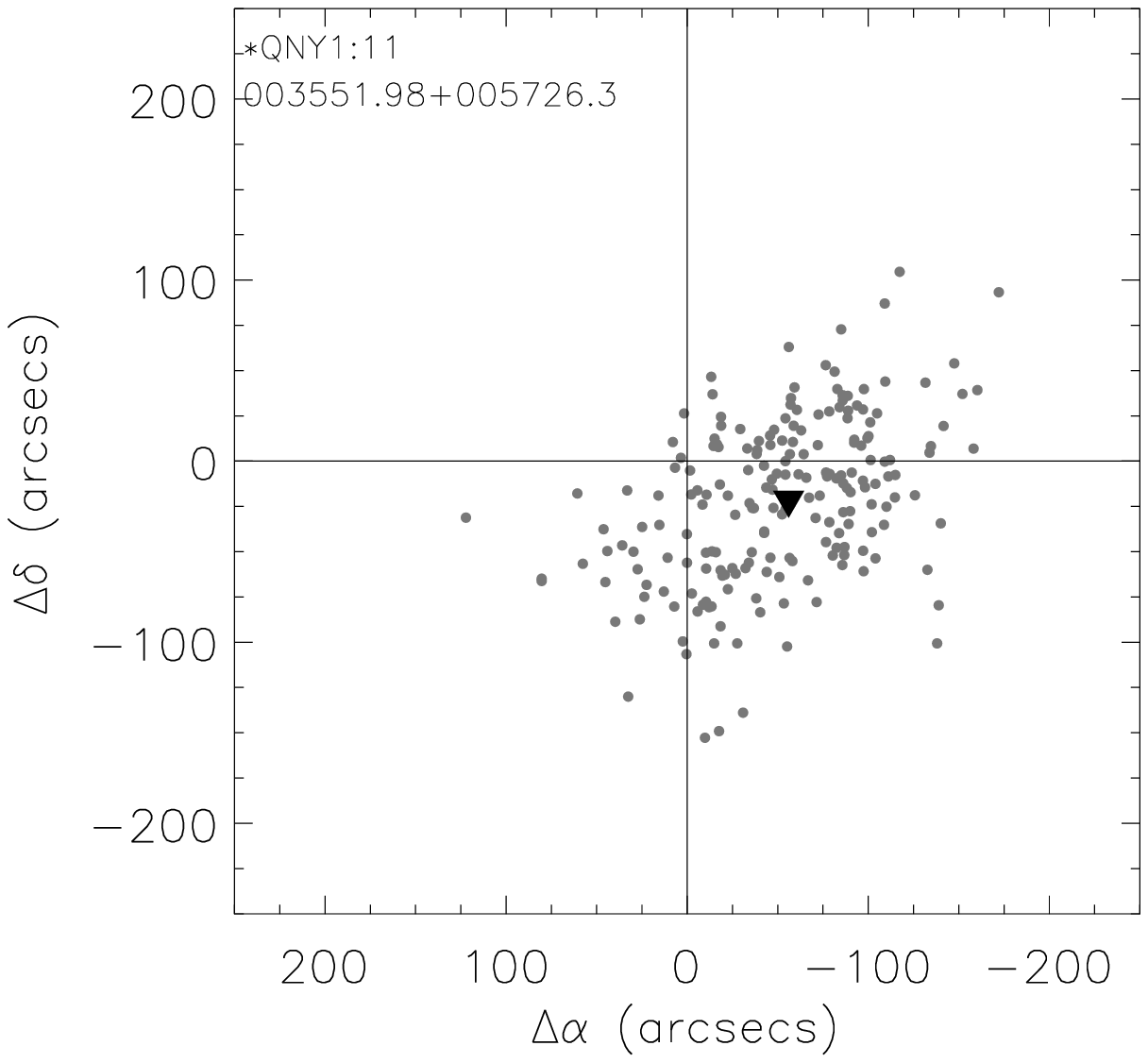}
\includegraphics[width=5.5cm]{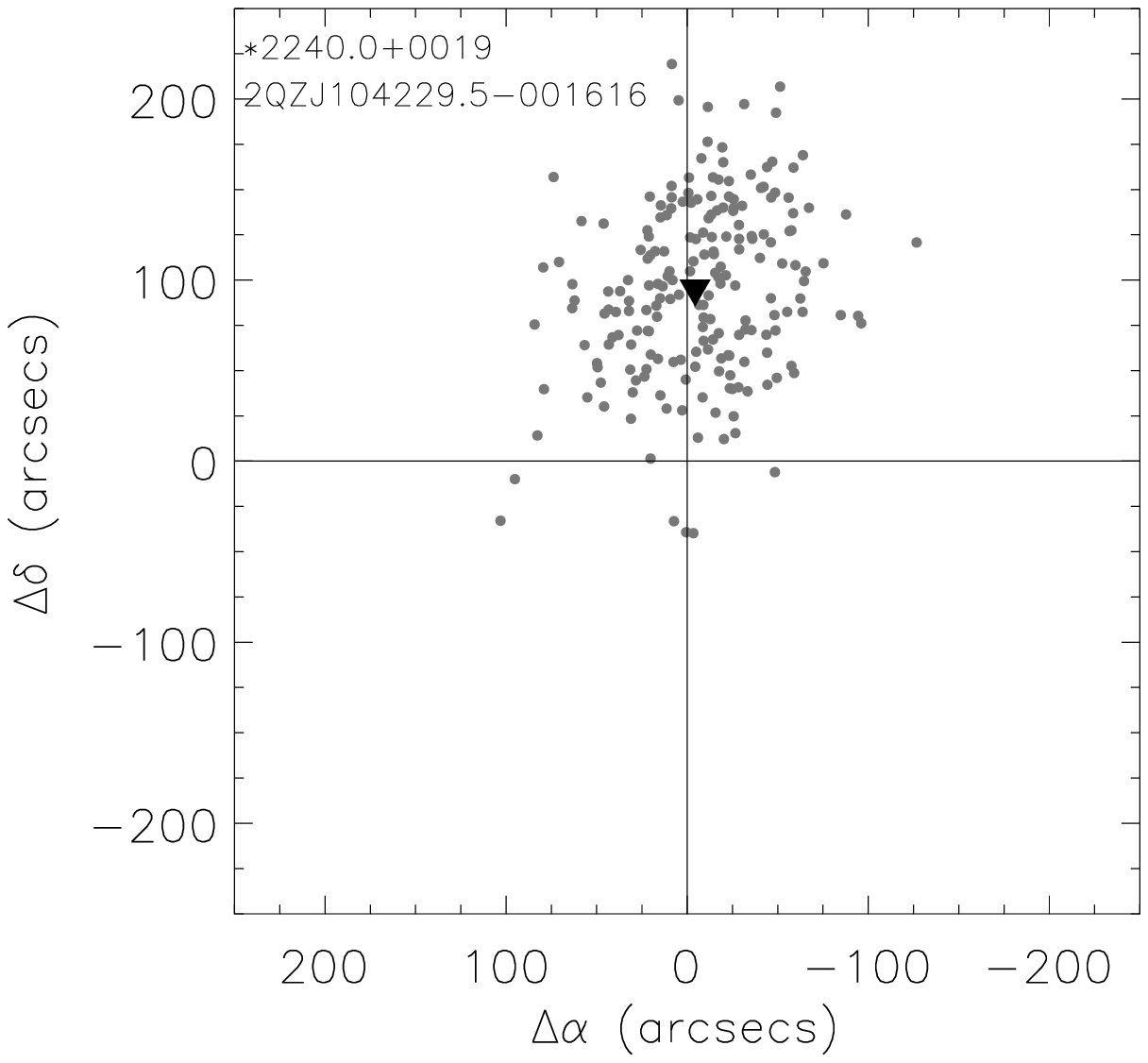}
\includegraphics[width=5.5cm]{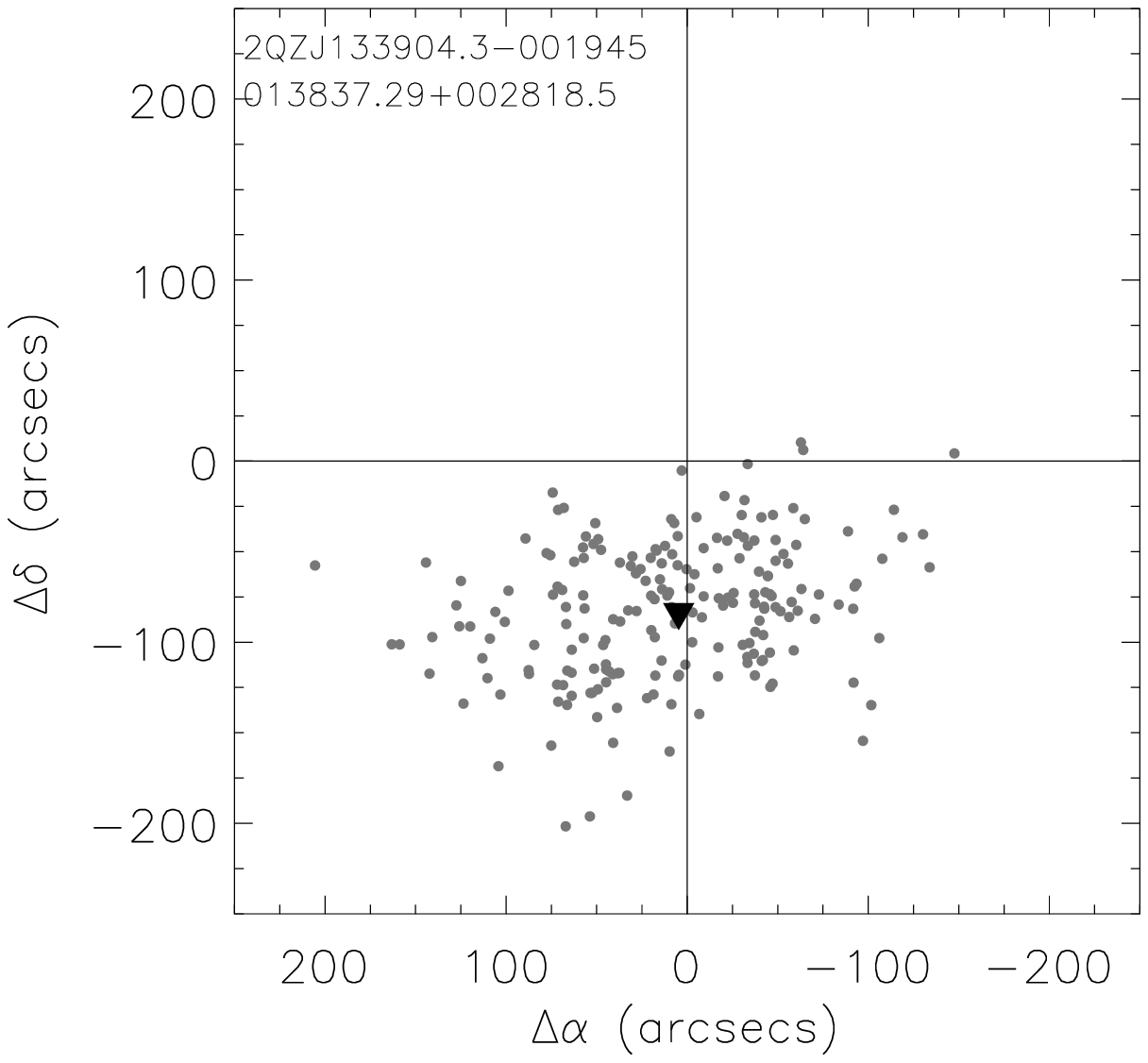}
\includegraphics[width=5.5cm]{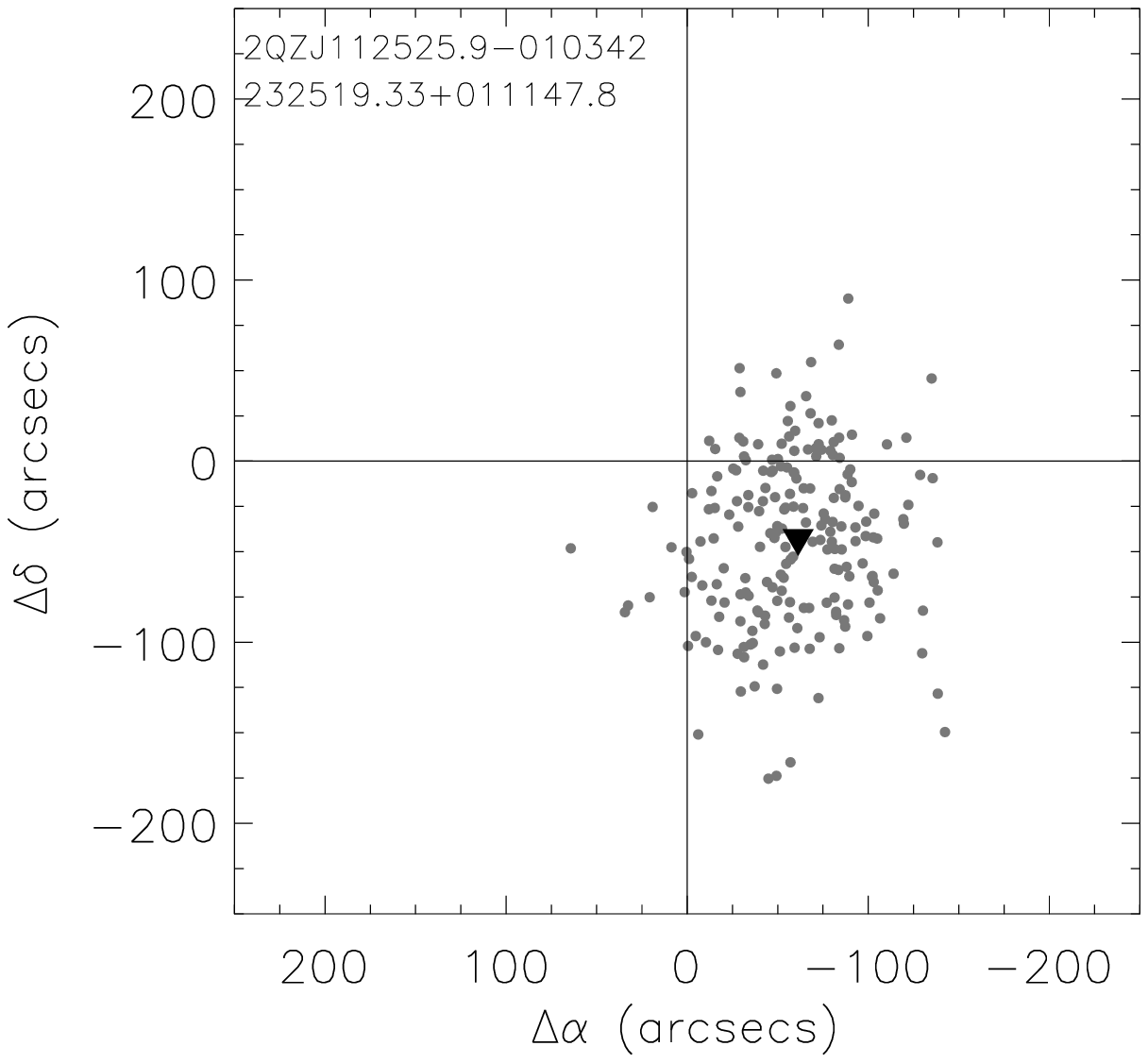}
\caption{Plots of the seven candidate ghost pairs indicating the
degree of anti-alignment, and its variation with the input
parameters. For each pair the cross is centred on the antipode of the first quasar,
 and the triangle marks the displacement of the second
quasar for the best values of the parameters
$\Omega_{\rm m}$, and $\bmath{v}_{\circ}$. In each case the second
quasar lies within the search radius of $134$ arcsec, which is how
they were selected. The dots show the computed displacements for the
Monte Carlo variations of the parameters $\Omega_{\rm m}$, and
$\bmath{v}_{\circ}$. This distribution overlaps the antipode in each
case, indicating that for possible values of the parameters
$\Omega_{\rm m}$, and $\bmath{v}_{\circ}$ the two quasars could be
perfectly anti-aligned.\label{fig:plots}}
\end{figure*}

The relative positions of the two quasars in each pair are plotted in
Figure \ref{fig:plots}. For each pair the plot is centred on the antipode of the first quasar 
in the pair,
and the points illustrate the offset of the second quasar from perfect
anti-alignment, for the 200 Monte Carlo realisations of the
parameters $\Omega_{\rm m}$, and $\bmath{v}_{\circ}$. In each case the
distribution of points overlaps perfect anti-alignment, indicating
that, given the uncertainties in the parameters, the pairs could be
real ghost images. As a further investigation, the angles between the
axes of each quasar pair were computed to search for any specific
angles that could indicate points on a lattice (such as $\pi/2$), but
nothing striking was discovered.  There are magnitude differences, up
to $\Delta m=2$, between the two quasars of a pair, but these might be
due to variability because of the epoch differences of the
observations, or time delay differences because the light paths to us
traverse different regions of space, or the differences could be
because each quasar is being viewed from a different direction. For
the same reasons it would not be unreasonable for the spectra to
differ also. A more important consideration is the number of pairs
that might be expected by chance.

\begin{table*}
\caption{Antipodal pairs of quasars within 134 arcsec and $\Delta z =
0.04$.  Redshifts preceded by \ddag\ are of lesser accuracy or
even wrong (see V\'eron catalogue for further details).}
\label{table:quasars}
\begin{centering}
\begin{scriptsize}
\begin{tabular}{lllllllll}
\hline
 Name & Position (J2000) &  & Corrected Position & & Redshift & Magnitude & Catalogue & Deviation from \\
 & R.A.& Dec & R.A. & Dec & & & & opposition\\
\hline
 %& & & & &  &\\
$\ast$Q 0007$+$0236 & 00h 10m 19.5s & $+$02\degr 53\arcmin 38\arcsec & 00h 10m 23.1s & $+$02\degr 49\arcmin 57.5\arcsec & 0.588 & V=18.4 & Veron & $\theta = 99\arcsec$\\
2QZJ121030.8-024456 & 12h 10m 30.88s & $-$02\degr 44\arcmin 56.3\arcsec & 12h 10m 27.2s & $-$02\degr 48\arcmin 39.7\arcsec & 0.6257 & b$_{\rm j}$=20.1 & 2dF & $\Delta z = 0.0377$\\
 & & & & & \\
$\ast$Q 0058$-$3424 & 01h 01m 14.1s & $-$34\degr 08\arcmin 31\arcsec & 01h 01m 34.7s & $-$34\degr 16\arcmin 58.0\arcsec & \ddag2.08 & V=19.36 & Veron & $\theta = 129\arcsec$\\
$\ast$KP 1259.4$+$34.7 & 13h 01m 46.5s & $+$34\degr 26\arcmin 34\arcsec & 13h 01m 25.8s & $+$34\degr 18\arcmin 06.0\arcsec & \ddag2.08 & V=18.5 & Veron & $\Delta z = 0.0000$\\
 & & & & & \\  
FIRST J1057$+$3119 & 10h 57m 5.2s & $+$31\degr 19\arcmin 07\arcsec & 10h 57m 06.7s & $+$31\degr 11\arcmin 01.5\arcsec & 1.332 & V=18.7 & Veron & $\theta = 27\arcsec$\\ 
2QZJ225706.5$-$310312 & 22h 57m 6.56s & $-$31\degr 03\arcmin 12.5\arcsec & 22h 57m 05.0s & $-$31\degr 11\arcmin 17.2\arcsec & 1.3398 & b$_{\rm j}$=20.1 & 2dF & $\Delta z = 0.0078$ \\
 & & & & & \\
$\ast$QNY1:11 & 12h 36m 9.9s & $-$00\degr 48\arcmin 03\arcsec & 12h 36m 02.8s & $-$00\degr 52\arcmin 56.0\arcsec & 1.874 & V=21.3 & Veron & $\theta = 60\arcsec$ \\
003551.98$+$005726.3 & 00h 35m 51.98s & $+$00\degr 57\arcmin 26.3\arcsec & 00h 35m 59.0s & $+$00\degr 52\arcmin 33.9\arcsec & 1.905 & g$\prime$=19.06 & SDSS & $\Delta z = 0.0310$\\
 & & & & & \\
$\ast$2240.0$+$0019 & 22h 42m 33.3s & $+$00\degr 27\arcmin 09\arcsec & 22h 42m 31.6s & $+$00\degr 22\arcmin 30.3\arcsec & \ddag2.0 & V=20.0 & Veron & $\theta = 95\arcsec$\\
2QZJ104229.5-001616 & 10h 42m 29.53s & $-$00\degr 16\arcmin 16.1\arcsec & 10h 42m 31.3s & $-$00\degr 20\arcmin 55.7\arcsec & 1.9713 & b$_{\rm j}$=19.7 & 2dF & $\Delta z = 0.0287$\\
 & & & & & \\
2QZJ133904.3$-$001945 & 13h 39m 4.35s & $-$00\degr 19\arcmin 45\arcsec & 13h 38m 50.6s & $-$00\degr 24\arcmin 44.3\arcsec & 1.3474 & b$_{\rm j}$=20.4 & 2dF & $\theta = 84\arcsec$\\
013837.29$+$002818.5 & 01h 38m 37.29s & $+$00\degr 28\arcmin 18.5\arcsec & 01h 38m 50.9s & $+$00\degr 23\arcmin 20.2\arcsec & 1.35 & g$\prime$=18.93 & SDSS & $\Delta z = 0.0026$\\
 & & & & & \\
2QZJ112525.9$-$010342 & 11h 25m 25.95s & $-$01\degr 03\arcmin 42.4\arcsec & 11h 25m 24.7s & $-$01\degr 08\arcmin 07.6\arcsec & 1.743 & b$_{\rm j}$=20.4 & 2dF & $\theta = 74\arcsec$ \\
232519.33$+$011147.8 & 23h 25m 19.33s & $+$01\degr 11\arcmin 47.8\arcsec & 23h 25m 20.6s & $+$01\degr 07\arcmin 24.4\arcsec & 1.724 & g$\prime$=18.45 & SDSS & $\Delta z = 0.0190$\\
\hline 
%& & & & & \\
\end{tabular}
\end{scriptsize}
\medskip
\end{centering}
\end{table*}

We investigated how the expected number of antipodal pairs varied with
respect to the search tolerance. We varied the search angular radius
from $\Theta = 0$ to $\Theta = 10$ arcmin, and the redshift
tolerance from $\Delta z = 0$ to $\Delta z = 0.4$. Although the
catalogue is inhomogeneous we found that the number of antipodal
quasar pairs recovered followed closely the relations
$\propto\Theta^2$ and $\propto\Delta z$ expected for unrelated
pairs. Therefore we used a fit to the relations at large $\Theta$,
$\Delta z$ to establish an estimate of the number of pairs expected by
chance.  For our search tolerance of $\Theta=134$ arcsec we would have
expected $12\pm3.5$ quasar pairs by chance. Our result of seven pairs,
therefore, is consistent with chance. While it remains possible that
one or more of the pairs could be a real ghost pair, this cannot be
proven from the existing data.

%%%%%%%%%%%%%%%%%%%%%%%%%%%%%%%%%%%%%%%%%%%%%%%%%%%%%%%%%%%%%%%%%
%%%%%%%%%%%%%%%%%%%%%%%%%%%%%%%%%%%%%%%%%%%%%%%%%%%%%%%%%%%%%%%%%

\subsection{Discussion}
We have presented a search for ghost images of our galaxy using the
method of antipodal quasar pair detection, correcting for the angular
distortion due to the history of motion of the local group. We have
found no significant difference in the number of pairs found over the
number we would expect purely by chance using our search tolerance on
this catalogue. We have examined the angles between the axis of each
quasar pair but failed to discover any indication of a lattice
structure.

The uncertainties both with our present knowledge of cosmological
parameters and the peculiar velocity of the local group make a search
such as this increasingly difficult as quasar catalogues increase in
size. To prove significance in a search such as this the search
tolerance needs to be reduced to a value where we would be expecting
only one or two detections purely by chance. In the future we can
expect improvements in the accuracy of the parameters
$\Omega_{\rm m}$, $\bmath{v}_{\circ}$, that will allow a smaller
search radius, that could compensate for the much larger future quasar
catalogues. However, soon the uncertain correction for the peculiar
motion of the Milky Way within the local group will become a
significant factor. Given the {\it COBE} constraints on the size of the
fundamental domain (and so the small possible number of ghost images),
and the relatively short lifetime of quasars, it would seem unlikely
that it will be possible to identify putative ghost images,
unambiguously, by this method. Instead, extension of the search to configurations
of three or more quasars on a lattice could yield a convincing detection.

%%%%%%%%%%%%%%%%%%%%%%%%%%%%%%%%%%%%%%%%%%%%%%%%%%%%%%%%%%%%%%%%%%
%%%%%%%%%%%%%%%%%%%%%%%%%%%%%%%%%%%%%%%%%%%%%%%%%%%%%%%%%%%%%%%%%%

\section*{Acknowledgments}
We are grateful to Andrew Jaffe, Thomas Babbedge, Matthew Pieri and Matthew Fox
for their helpful comments. The 2dF QSO Redshift Survey (2QZ) was 
compiled by the 2QZ survey team from observations made with the 2-degree Field on the Anglo-Australian
Telescope. Funding for the creation and distribution of the SDSS Archive has been provided by the Alfred P. Sloan Foundation, the Participating Institutions, NASA, NSF, the U.S. Department of Energy, the Japanese Monbukagakusho, and the Max Planck Society. The SDSS website is http://www.sdss.org.

%%%%%%%%%%%%%%%%%%%%%%%%%%%%%%%%%%%%%%%%%%%%%%%%%%%%%%%%%%%%%%%%%%
%%%%%%%%%%%%%%%%%%%%%%%%%%%%%%%%%%%%%%%%%%%%%%%%%%%%%%%%%%%%%%%%%%

\bsp

\label{lastpage}


\begin{thebibliography}{}

\bibitem[\protect\citeauthoryear{Bond, Pogosyan \& Souradeep}{2000}]{Bond}
Bond J. R., Pogosyan D. and Souradeep T., 2000, PhysRevD, 62, 043006

\bibitem[\protect\citeauthoryear{Croom et al.}{2001}]{Croom}
Croom S.ÊM., Smith R.ÊJ., Boyle B.ÊJ. Shanks T., Loaring N.ÊS., Miller L., Lewis I.ÊJ., 2001, MNRAS, 322, L29

\bibitem[\protect\citeauthoryear{Fagundes \& Wichoski}{1987}]{F1987}
Fagundes H. V., Wichoski U. F., 1987, ApJ, 322, L5

\bibitem[\protect\citeauthoryear{Fagundes}{1989}]{F1989}
Fagundes H. V., 1989, ApJ, 338, 618

\bibitem[\protect\citeauthoryear{Kogut et al.}{1993}]{Kogut}
Kogut A. et al., 1993, ApJ, 419, 1

\bibitem[\protect\citeauthoryear{Lachi\`eze--Rey \& Luminet}{1995}]{LRL}
Lachi\`eze--Rey M., Luminet J.--P., 1995, Phys. Rep., 254, 135

\bibitem[\protect\citeauthoryear{Lahav et al.}{1991}]{Lahav}
Lahav O., Lilje P. B., Primack J.R., Rees M.J., 1991, MNRAS, 251, 128

\bibitem[\protect\citeauthoryear{Levin}{1998}]{L1998}
Levin J., Scannapieco E., Silk J., 1998, PhysRevD, 58, 103516

\bibitem[\protect\citeauthoryear{Peebles}{1993}]{P1993}
Peebles P. J. E., 1993, Principles of Physical Cosmology. Princeton University Press, Princeton, New Jersey.

\bibitem[\protect\citeauthoryear{Roukema}{1996}]{R1996}
Roukema B. F., 1996, MNRAS, 283, 1147

\bibitem[\protect\citeauthoryear{Schneider et al.}{2001}]{SDSS}
Schneider D. P. et al., 2002, AJ, 123, 567

\bibitem[\protect\citeauthoryear{Verde et al.}{2002}]{Verde}
Verde L., et al., 2002, MNRAS, 335, 432

\bibitem[\protect\citeauthoryear{V\'eron-Cetty \& V\'eron}{2002}]{VV}
V\'eron-Cetty M.-P., V\'eron P., 2001, A Catalogue of Quasars and Active Nuclei, 10th edition, Observatoire de Haute Provence, CNRS, F-04870 Saint-Michel l'Observatoire. 


\end{thebibliography}
\end{document}